 \documentclass[journal, comsoc]{IEEEtran}

\usepackage{gensymb}
\usepackage{cite}
\usepackage{amsmath,amssymb,amsfonts}
\usepackage{graphicx}
\usepackage{xcolor}
\usepackage{kotex}  
\usepackage{caption}
\usepackage{subcaption}
\usepackage{caption}
\usepackage{subcaption}
\usepackage{tabularx}
\usepackage{booktabs}

\usepackage{stfloats}
\usepackage{float}
\usepackage{wrapfig}
\usepackage{xcolor}
\usepackage[table]{xcolor}
\usepackage[table,svgnames]{xcolor}
\usepackage[dvipsnames]{xcolor}
\usepackage{colortbl} 
\usepackage{color,soul}
\usepackage{url}
\usepackage{verbatim}
\usepackage{graphicx}
\usepackage{cite}
\usepackage{adjustbox}

\usepackage[acronym,toc,shortcuts]{glossaries}
\newacronym{BS}{BS}{Base Stations}
\newacronym{NTN}{NTN}{Non-Terrestrial Networks}
\newacronym{TN}{TN}{Terrestrial Networks}

\newacronym{NOMA}{NOMA}{Non-Orthogonal Multiple Access}
\newacronym{IoT}{IoT}{Internet of Things}
\newacronym{UEs}{UEs}{User Equipment's}
\newacronym{UE}{UE}{User Equipment}
\newacronym{UAV}{UAV}{Uncrewed Aerial Vehicles}
\newacronym{SDG}{SDG}{Sustainable Development Goals}
\newacronym{UN}{UN}{United Nations}
\newacronym{HAPS}{HAPS}{High-Altitude Platform Stations}
\newacronym{AI}{AI}{Artificial Intelligence}
\newacronym{EE}{EE}{Energy Efficiency}
\newacronym{3GPP}{3GPP}{3$^{\text{rd}}$ Generation Partnership Project}
\newacronym{THz}{THz}{Terahertz}
\newacronym{mmWave}{mmWave}{millimeter Wave}
\newacronym{PLMN}{PLMN}{Public Land Mobile Networks}
\newacronym{MNO}{MNO}{Mobile Network Operators}
\newacronym{LEO}{LEO}{Low Earth Orbit}
\newacronym{KPI}{KPI}{Key Performance Indicators}
\newacronym{GEO}{GEO}{Geostationary Equatorial Orbit}
\newacronym{MEO}{MEO}{Medium-Earth Orbit}
\newacronym{CapEx}{CapEx}{Capital Expenditures}

\newacronym{SA}{SA}{System and Architecture}
\newacronym{RAN}{RAN}{Radio Access Network}
\newacronym{NGSO}{NGSO}{Non-Geostationary Satellite Orbit}
\newacronym{UPF}{UPF}{User Plane Function}
\newacronym{mMIMO}{mMIMO}{massive Multiple Input Multiple Output}
\newacronym{QoS}{QoS}{Quality-of-Service}
\newacronym{LAN}{LAN}{Local Area Network}
\newacronym{RHS}{RHS}{Reconfigurable Holographic Surfaces}
\newacronym{ISAC}{ISAC}{Integrated Sensing and Communication}
\newacronym{SNO}{SNO}{Satellite Network Operators}
\newacronym{MENA}{MENA}{Middle East and North Africa}
\newacronym{LTE}{LTE}{Long-Term Evolution}
\newacronym{NAS}{NAS}{Non Access Stratum}
\newacronym{Wi-Fi}{Wi-Fi}{Wireless Fidelity}
\newacronym{FFR}{FFR}{Full Frequency Reuse}
\newacronym{ZF}{ZF}{Zero Forcing}
\newacronym{MIMO}{MIMO}{Multiple-Input-Multiple-Output}
\newacronym{MMSE}{MMSE}{Minimum Mean Square Error}
\newacronym{RAT}{RAT}{Radio Access Technologies}
\newacronym{CSI}{CSI}{Channel State Information}
\newacronym{CQI}{CQI}{Channel Quality Indicator}
\newacronym{RS}{RS}{Reference Signal}
\newacronym{GNSS}{GNSS}{Global Navigation Satellite System}
\newacronym{AMF}{AMF}{Access and Mobility Management Function}
\newacronym{LMF}{LMF}{Location Management Function}
\newacronym{5GC}{5GC}{5G Core Network}
\newacronym{RF}{RF}{Radio Frequency}
\newacronym{NC-JT}{NC-JT}{Non-Coherent JT}
\newacronym{MC}{MC}{Multi-Connectivity}
\newacronym{RRM}{RRM}{Radio Resource Management}
\newacronym{ML}{ML}{Machine Learning}
\newacronym{RU}{RU}{Radio Unit}
\newacronym{CU}{CU}{Control Unit}
\newacronym{DU}{DU}{Distributed Unit}
\newacronym{RIC}{RIC}{RAN Intelligent Controller}
\newacronym{CN}{CN}{Core Network}
\newacronym{EIRP}{EIRP}{Equivalent Isotropic Radiated Power}
\newacronym{VSAT}{VSAT}{Very Small Aperture Terminal}
\newacronym{RTT}{RTT}{Round-Trip Time}
\newacronym{RIS}{RIS}{Reconfigurable Intelligent Surfaces}
\newacronym{CR}{CR}{Cognitive Radios}
\newacronym{ITU}{ITU}{International Telecommunication Union}
\newacronym{gNB}{gNB}{Next-Generation Node B}
\newacronym{6G}{6G}{Sixth Generation}
\newacronym{5G}{5G}{Fifth Generation}
\newacronym{4G}{4G}{Fourth Generation}
\newacronym{OPEX}{OPEX}{Operational Expenditure}
\newacronym{FL}{FL}{Federated Learning}
\newacronym{AP}{AP}{Access Points}

\usepackage{multirow}
\usepackage{array,boldline, makecell,booktabs}
\usepackage{longtable} 

    \usepackage[group-separator={,},group-minimum-digits=4]{siunitx}

\begin{document}
%
\title{Net-Zero 6G from Earth to Orbit: Sustainable Design of Integrated Terrestrial and Non-Terrestrial Networks}
%
%
%

\author{Muhammad Ali Jamshed,~\IEEEmembership{Senior Member,~IEEE}, Malik Muhammad Saad, Muhammad Ahmed Mohsin,~\IEEEmembership{Graduate Member,~IEEE,}, Dongkyun Kim,~\IEEEmembership{Member,~IEEE}, Octavia A. Dobre~\IEEEmembership{Fellow,~IEEE}, Halim Yanikomeroglu, ~\IEEEmembership{Fellow,~IEEE,}  Lina Mohjazi, ~\IEEEmembership{Senior Member,~IEEE}.
\thanks{M. A. Jamshed is with the College of Science and Engineering, University	of Glasgow, UK (e-mail: muhammadali.jamshed@glasgow.ac.uk).\\ 
$~~~$M. M. Saad is with the School of Computer Science and Engineering, Kyungpook National University (KNU), Daegu, Korea (email: maliksaad@knu.ac.kr).\\
$~~~$M. A. Mohsin is with the Department of Electrical Engineering, Stanford University, Stanford, CA, USA (email: muahmed@stanford.edu).\\
$~~~$D. Kim is with the School of Computer Science and Engineering, Kyungpook National University (KNU), Daegu, Korea (email: dongkyun@knu.ac.kr).\\
$~~~$O. A. Dobre is with the Faculty of Engineering and Applied Science, Memorial University, Canada (email: odobre@mun.ca)\\
$~~~$H. Yanikomeroglu is with the Department of Systems and Computer Engineering at Carleton University, Ottawa, Canada (e-mail: halim@sce.carleton.ca)
$~~~$L. Mohjazi is with the James Watt School of Engineering, University	of Glasgow, UK (e-mail: lina.mohjazi@glasgow.ac.uk).\\ 
}
}

%
%

\markboth{Submitted to IEEE Communication Magazine}%
{Shell \MakeLowercase{\textit{et al.}}: Bare Demo of IEEEtran.cls for IEEE Journals}

\maketitle

\begin{abstract}
The integration of Terrestrial Networks (TN) and Non-Terrestrial Networks (NTN) plays a crucial role in bridging the digital divide and enabling Sixth Generation (6G) and beyond to achieve truly ubiquitous connectivity. However, combining TN and NTN introduces significant energy challenges due to the diverse characteristics and operational environments of these systems. In this paper, we present for the first time a comprehensive overview of the design challenges associated with achieving Net-Zero energy targets in integrated TN and NTN systems. We outline a set of key enabling technologies that can support the energy demands of such networks while aligning with Net-Zero objectives. To enhance the Energy Efficiency (EE) of integrated TN and NTN systems, we provide a use case analysis that leverages Artificial Intelligence (AI) to deliver adaptable solutions across diverse deployment scenarios. Finally, we highlight promising research directions that can guide the sustainable evolution of integrated TN and NTN.
\end{abstract}


\IEEEpeerreviewmaketitle


\section{Introduction} 

The exponential growth of mobile data traffic is placing increasing pressure on the energy demands of wireless networks. Terrestrial \ac{BS} already account for the majority of the total energy consumption of the network. To address this, strategies such as cell switching, which deactivate underutilized \ac{BS} during periods of low traffic, have proven effective in \ac{TN}. However, these approaches face inherent limitations due to the coverage and capacity constraints of terrestrial infrastructure alone, underscoring the need for \ac{NTN} integration. The integration of \ac{TN} and \ac{NTN} is gaining traction as a promising solution to extend connectivity to remote, underserved, and disaster-affected regions. Many \ac{NTN} platforms are powered by renewable energy sources, such as solar power, improving sustainability and reducing dependence on grid electricity. Furthermore, \ac{TN} and \ac{NTN} integration enables dynamic traffic management, flexible load balancing and rapid deployment in emergencies, aligned with the long-term goals of the \ac{ITU} of universal connectivity and environmental sustainability \cite{bariah2022ris}.

Incorporating \ac{NTN} elements such as, satellites, \ac{HAPS} and \ac{UAV}, into \ac{TN} systems offers several advantages. They complement \ac{TN} by providing additional coverage and capacity, thereby reducing reliance on energy-intensive terrestrial \ac{BS}. In an integrated \ac{TN} and \ac{NTN} environment, however, user devices must support multiple \ac{RAT} to connect across both domains \cite{11122487}. While this enhances reliability and service continuity, it also increases the overall density of wireless infrastructure, which can raise total energy consumption and associated carbon emissions. \ac{EE} therefore becomes a critical concern. \ac{NTN} platforms often operate with strict power limitations in challenging environments where energy replenishment is expensive and logistically complex \cite{mohsin2025hierarchicaldeepreinforcementlearning}. Advanced energy management strategies, combined with the adoption of renewable energy sources, are essential to extend the lifetime of \ac{NTN} components, reduce \ac{OPEX}, and minimize environmental impact. Ensuring \ac{EE} in integrated \ac{TN} and \ac{NTN} is thus both a necessity and an urgent priority. 

This article, for the first time and to the best of our knowledge, provides a comprehensive overview of the design challenges involved in achieving Net-Zero targets for integrated \ac{TN} and \ac{NTN} systems. In addition, it identifies the main enabling technologies, reviews current solutions for energy-efficient operation, and discusses potential directions towards sustainable and resilient \ac{6G} networks. Moreover, the contribution to the knowledge of this article is summarized as follows:

\begin{itemize}
    \item For an integrated \ac{TN} and \ac{NTN} system, we provide an extensive overview of the design challenges associated with achieving Net-Zero targets.
    \item We present a thorough analysis of the technologies that improve the energy requirements of integrated \ac{TN} and \ac{NTN}.
    \item We introduce a use case demonstration of how \ac{EE} of integrated \ac{TN} and \ac{NTN} could be improved by utilizing \ac{AI}.
    \item We provide a future roadmap of technologies that could shape the sustainable evolution of integrated \ac{TN} and \ac{NTN}.
\end{itemize}

The rest of the article is organized as follows. Section II discusses the background and motivation for integrating \ac{TN} and \ac{NTN}, and underscores the importance of maintaining alignment with the Net-Zero targets. In an integrated \ac{TN} and \ac{NTN} environment, Section III reviews the key design challenges impeding Net-Zero goals. Section IV outlines the primary technological and operational enablers facilitating the transition toward carbon-neutral integrated \ac{TN} and \ac{NTN}. Section V demonstrates a use case analysis of improving the \ac{EE} of an integrated \ac{TN} and \ac{NTN} communication scenario using a proposed algorithm that jointly utilizes \ac{AI} and \ac{NOMA}. Section V provides an overview of future research challenges, and finally, Section VII concludes the article.

\section{Background and Motivation}
The \ac{ITU} International Mobile Telecommunication (IMT)-2030 framework identifies enhanced coverage and sustainability as the core objectives for \ac{6G}. In parallel, the \ac{3GPP} standardization has matured the technical basis for integrating \ac{TN} and \ac{NTN}, so that a single device can seamlessly access both ground and space segments and promises to close long-standing coverage gaps across sparsely populated regions. Meeting these coverage and sustainability goals cannot come at the expense of rising energy use and emissions. \ac{RAN} already accounts for an approximately 70\% share of the end-to-end network power consumption, with energy drawn even during low-traffic periods. Left unaddressed, densification on the ground to meet capacity demand and the addition of space/air segments for coverage could increase the absolute energy use and \ac{OPEX}. Conversely, \ac{NTN} elements operate under strict power budgets: satellites rely on solar arrays and batteries subject to eclipse cycles, while \ac{HAPS} must balance payload mass, aerodynamics, and available solar energy. Coordinating these heterogeneous constraints across the ground, air, and space makes energy a first-class system design concern rather than a post-deployment optimization \cite{11003502}.

Integrated \ac{TN} and \ac{NTN} must reconcile very different propagation and protocol regimes. Terrestrial paths can deliver millisecond (ms)-level delay for real-time applications; by contrast, space links introduce an additional propagation delay that varies with altitude and orbit. Empirical and operator-reported measurements place typical \ac{RTT} at tens of ms for \ac{LEO}, roughly one to a few hundred ms for \ac{MEO}, and on the order of 600--700~ms for \ac{GEO}. Long and variable \ac{RTT} stress conventional retransmission and control procedures, and increase the energy cost of na\"ive redundancy. Likewise, extremely large satellite beam footprints enable wide-area reach but limit spatial frequency reuse compared to small terrestrial cells, creating tension between wide coverage and per-user capacity. Finally, frequent handovers caused by fast-moving \ac{LEO} satellites, Doppler shifts in the kilohertz (kHz) range, and discontinuous contact windows add control overheads that, if unmanaged, translate into wasted energy at both the device and the network.

\ac{3GPP} Release~17 delivered the first normative \ac{NTN} specifications, covering deployment scenarios, architecture options and fundamental radio procedures. Release~18 (\ac{5G}-Advanced) elevates \ac{EE} and intelligent automation as priority areas throughout the system, laying the foundations for coordinated and energy-aware operation in heterogeneous topologies that include space/air segments \cite{10847914}. In parallel, industry case studies demonstrate that substantial site-level energy savings are achievable with modern radio hardware, software features, and traffic-aware sleep with reported improvements ranging from roughly 50\% to 70\% under field conditions. It is evidence that the energy reduction trajectory can be extended to integrated \ac{TN} and \ac{NTN} if designed from the outset. Together, these developments motivate a focused treatment of \emph{Net-Zero by design} for integrated \ac{TN} and \ac{NTN}.

The research gap is twofold. First, there is a need for a consolidated view of \emph{where} energy is spent across ground, air, and space segments, and \emph{how} design decisions (coverage partitioning, spectrum reuse, control-plane procedures, backhaul placement, and timing) propagate into energy and carbon outcomes. Second, there is a need for validated, system-level enablers (renewable-powered sites and gateways, energy-cost-aware scheduling, passive and reconfigurable surfaces, and \ac{ML}-driven orchestration) that reduce energy per delivered bit while preserving \ac{QoS}. This article addresses these gaps by (i) synthesizing the design challenges that uniquely arise in \ac{TN} and \ac{NTN} integration from a Net-Zero perspective, (ii) cataloging the enabling technologies and operational practices aligned with current standards, and (iii) presenting a use case that leverages \ac{AI} and access-layer techniques to improve \ac{EE} without degrading coverage or reliability.

\section{Net-Zero Design Challenges for Integrated TN and NTN}
Integrated \ac{TN} and \ac{NTN} promises to bridge the digital divide through ubiquitous coverage, but doing so sustainably entails numerous technical and operational challenges. In essence, \ac{TN} and \ac{NTN} integration must dramatically expand coverage and capacity while curbing energy consumption and carbon footprint. This section surveys the key design challenges impeding Net-Zero goals in integrated \ac{TN} and \ac{NTN}, drawing on recent literature and standardization efforts to highlight each issue. Each challenge is coupled with potential directions or considerations from current research, ensuring that the discussion remains factual and grounded in contemporary work. Moreover, a comparative analysis of these design challenges is provided in Table \ref{tab:tn-ntn-challenges}.

\begin{table*}[!t]
\centering
\caption{Design challenges and Net-Zero enablers across TN, NTN, and integrated TN and NTN.}
\label{tab:tn-ntn-challenges}
\scriptsize
\setlength{\tabcolsep}{4pt}
\renewcommand{\arraystretch}{1.15}
\begin{tabularx}{\textwidth}{|X|X|X|X|X|}
\toprule
\rowcolor{LightSteelBlue}
\textbf{Challenge area} &
\textbf{TN: Core Challenges \& Fix} &
\textbf{NTN: Core Challenges \& Fix} &
\textbf{Integrated TN \& NTN: Joint Challenges} &
\textbf{Net-Zero Oriented Enablers} \\
\midrule
\rowcolor{AliceBlue}
Coverage \& footprint &
Rural sites are expensive and often under-utilized, use temporary cells, shared sites, and demand-based activation. &
Very wide beams give reach but little spatial reuse, use multiple spot beams, beam hopping, and diverse backhaul. &
Decide which layer serves which users without wasting energy in either domain. &
Carbon-aware coverage planning; serve sparse regions from space/air, keep dense traffic on the ground. \\
\midrule
\rowcolor{AliceBlue}
Latency \& jitter &
A ms-level delay; spikes from congested backhaul, place compute and caches near users and schedule by \ac{QoS}. &
Propagation through space adds tens to hundreds of ms, cache content and tune protocols for long paths. &
Keep time-critical traffic on the ground; offload delay-tolerant flows to satellite or \ac{HAPS}. &
Policy-based path selection; transport tuned for mixed delays. \\
\midrule
\rowcolor{AliceBlue}
Energy \& power budget &
\ac{BS} draw high power even when idle, deep sleep, smarter cooling, and antenna muting. &
Space/air platforms have strict solar-battery limits, duty cycling and power-aware waveforms. &
End-to-end energy budgeting across device, ground, and space; avoid dual-connectivity waste. &
Schedulers that price energy; renewable-powered sites and gateways; carbon-aware control. \\
\midrule
\rowcolor{AliceBlue}
Spectrum \& interference &
Dense reuse creates inter-cell interference, directional beams and coordinated scheduling. &
Adjacent beams and gateway links can interfere, beam shaping and careful frequency plans. &
Coexistence when both layers use the same bands and locations. &
Dynamic spectrum sharing; passive Reconfigurable Intelligent Surfaces (RIS); broadcast for common content. \\
\midrule 
\rowcolor{AliceBlue}
Propagation \& link budget &
Blockage and indoor loss at high frequencies, use repeaters and fall back to lower bands when needed. &
Free-space loss and rain fade on space links, site diversity and fade mitigation. &
Mix of very different channels across domains. &
Hybrid free-space-optical/radio backhaul; diversity combining on critical links. \\
\midrule
\rowcolor{AliceBlue}
Mobility \& handover &
Cell-edge ping-pong, make-before-break and parameter tuning. &
Short satellite passes cause frequent handovers, predictive tracking and beam pre-selection. &
Session continuity across ground, air, and space without drops. &
Multi-connectivity and buffer-assisted handover. \\
\midrule
\rowcolor{AliceBlue}
Doppler \& synchronization &
Vehicular motion distorts frequency at millimeter-wave, robust pilots and short transmission times. &
Fast-moving satellites create large Doppler shift and variable delay, pre-compensation and tolerant numerologies. &
Frequent re-synchronization burdens devices and schedulers. &
Assistance data (including satellite ephemeris) and accurate time transfer. \\
\midrule
\rowcolor{AliceBlue}
Backhaul \& compute placement &
Rural fiber is scarce; high-power microwave links, use E-band or optical wireless. &
Gateway bottlenecks and feeder-link limits, inter-satellite mesh and diverse gateways. &
Where to anchor user-plane functions and caches for least energy and delay. &
Place compute near gateways; steer flows between core and gateways. \\
\midrule
\rowcolor{AliceBlue}
Orchestration \& automation &
Policy drift across vendors, closed-loop automation with clear intents. &
Limited telemetry and slow control loops, more onboard autonomy and scheduled reconfiguration. &
Avoid conflicting ground/space policies while meeting service targets. &
Energy-aware automation and digital twins for what-if planning. \\
\midrule
\rowcolor{AliceBlue}
Device constraints &
Network search drains batteries; limited antenna size, efficient discovery and extended sleep. &
High-gain antennas and search energy, assisted discovery and narrow beams. &
Two radios increase cost, heat, and idle power. &
Shared radio front-end; power-aware traffic steering; assisted satellite discovery. \\
\midrule
\rowcolor{AliceBlue}
Protocol \& air interface &
Heavy control signaling and channel estimation, lean carriers and fewer reference signals. &
Standard retransmission logic breaks with long delays, larger windows or fewer acknowledgments. &
Converged timing and robust measurements across domains. &
Standardized satellite access features; transport protocols tuned for mixed paths. \\
\midrule
\rowcolor{AliceBlue}
Policy, standards, \& carbon accountability &
Licensing and site rules vary by country, shared infrastructure and refarming where possible. &
International filings and interference limits, global coordination across orbits and beams. &
Roaming and billing across both domains with clear responsibilities. &
Energy and carbon metrics in contracts; hybrid licensing aligned with Net-Zero goals. \\
\midrule
\rowcolor{AliceBlue}
Security \& resilience &
Grid outages and vandalism, hardening, microgrids, and redundant paths. &
Jamming and space weather, redundancy, shielding, and safe-modes. &
End-to-end trust and key management across domains. &
Path diversity and modern zero-trust practices. \\
\midrule
\rowcolor{AliceBlue}
Economics \& business &
High rural cost with low revenue, infrastructure sharing and open interfaces. &
Launch and gateway costs; narrow markets, wholesale and neutral-host models. &
Settlements and partner operations across layers. &
Carbon pricing, green finance, and energy-based service targets. \\
\midrule
\rowcolor{AliceBlue}
Sustainability metrics \& Key Performance Indicators (KPI) &
Site-level energy often opaque, metering with radio-access energy metrics. &
Payload energy budgets vary with eclipse and orbit, mission energy dashboards. &
Whole-network life-cycle accounting across domains. &
Carbon-per-bit targets and telemetry fed into automation systems. \\
\bottomrule
\end{tabularx}
\end{table*}


\subsection{Coverage vs. Energy Efficiency} One fundamental challenge is achieving wide-area coverage without a proportional surge in energy use and emissions. \ac{TN} traditionally rely on dense deployments of \ac{BS} to extend coverage, which can be unsustainable in low-density or remote regions. In such areas, installing numerous towers yields very low utilization ($\leq$ 1\% in extreme cases) yet incurs high power consumption and maintenance costs. Each new \ac{5G} \ac{BS} can consume roughly three times the power of a \ac{4G} site (plus an extra ~1 kW per sector for \ac{mMIMO} antennas). Thus, purely terrestrial coverage expansion would significantly increase the network’s carbon footprint.

\ac{NTN}, by contrast, can blanket vast areas, bringing connectivity to rural hinterlands, oceans, and skies. This broad coverage comes with a sustainability advantage: offloading sparsely populated areas to \ac{NTN} avoids continuously powering underutilized \ac{BS}, thereby reducing the carbon footprint of ubiquitous connectivity. However, this solution introduces a coverage–quality trade-off. The very attribute that makes \ac{NTN} attractive for coverage, their large footprint, also means limited capacity and higher latency per user. Designing Net-Zero integrated \ac{TN} and \ac{NTN} thus requires balancing wide coverage with efficient capacity provisioning. Researchers emphasize that hybrid architectures must intelligently assign coverage responsibilities: latency-sensitive or high-throughput services should be handled by terrestrial cells when available, while \ac{NTN} serve as a gap filler for basic connectivity. This dynamic offloading can ensure broad coverage and sustainability, but it demands \ac{AI}-based network management.



\subsection{Latency and Signal Propagation vs. Energy Efficiency}
Integrating \ac{TN} and \ac{NTN} also forces engineers to reconcile drastically different latency regimes. Terrestrial links typically span tens of kilometers (km) at most, enabling \ac{RTT} latencies on the order of a few ms. In contrast, satellite communications introduce substantial propagation delays due to the distances involved. For example, a \ac{GEO} satellite at ~36,000 km altitude incurs about 600 ms \ac{RTT} latency, a delay unacceptable for real-time applications like interactive gaming or industrial control. Even \ac{LEO} satellites at ~500–1200 km, still have noticeable latency and rapid motion relative to Earth. \ac{HAPS} and \ac{UAV} flying at 8–20 km altitude introduce less delay than satellites but still more than terrestrial fiber or microwave links.

Elevated latency in \ac{NTN} is a critical design aspect; therefore, integrated \ac{TN} and \ac{NTN} systems must intelligently route traffic: time-sensitive data should remain on terrestrial paths whenever possible, whereas delay-tolerant traffic can be offloaded to satellites. Hybrid networking strategies are being investigated where \ac{NTN} serves as a backup or broadcast channel, while the \ac{TN} handles real-time communications. Despite such strategies, ensuring seamless user experience across heterogeneous latency links remains a challenge. Variations in the propagation delay across a satellite’s coverage complicate synchronization and scheduling. Closing the latency gap is not only a technical hurdle, but also an energy one: overcoming delay via error correction, retransmissions, or backup links can increase energy use. A Net-Zero design must find latency mitigation techniques that are energy-efficient, avoiding brute-force solutions (such as excessively redundant transmissions) that would waste power.

\subsection{Power Constraints vs. Energy Efficiency}
The most pivotal Net-Zero challenge is managing energy consumption in integrated \ac{TN} and \ac{NTN}. \ac{5G} \ac{BS} consume significantly more power than \ac{4G} ones, potentially 4–5 times higher total energy consumption per site. \ac{NTN} come with the opposite problem, i.e. energy scarcity. Satellites, \ac{HAPS}, and \ac{UAV} operate under strict power limitations, usually relying on solar panels and batteries onboard. When \ac{TN} and \ac{NTN} are combined, these issues intersect. Integrated networks can potentially save energy by leveraging \ac{NTN} for coverage, but they also introduce new energy overhead in coordination between domains. Consider a scenario of a smartphone that switches between terrestrial and satellite connectivity. If not optimized, the device might consume extra power constantly searching for satellite signals or maintaining dual connectivity. Likewise, satellite gateways on the ground must interface with terrestrial core networks, adding conversion and processing steps that consume power. Resource management becomes a 3D problem: balancing energy use between ground infrastructure and space/aerial infrastructure.

\ac{3GPP} has begun developing \ac{EE} enhancements specifically for \ac{NTN}. One approach is introducing an energy cost function into network scheduling algorithms to factor in the energy expense of using a satellite or \ac{UAV} link. By assigning an energy price to various resources, the network can dynamically determine the most energy-efficient approach for data delivery. Additionally, advanced sleep modes and adaptive duty cycling are proposed for \ac{NTN} components: for instance, a satellite or \ac{HAPS} can power down transponders when user demand is low. \ac{AI} algorithms for intelligent resource allocation can predict traffic and turn off or scale down power to certain \ac{NTN} nodes. Such measures are aimed at reducing the energy footprint of integrated \ac{TN} and \ac{NTN} without degrading \ac{QoS}. Furthermore, integrating renewable energy sources on the ground and in space is crucial, e.g., solar-powered \ac{BS} and highly efficient photovoltaic systems in satellites.

\subsection{Capacity vs. Energy Efficiency}
Efficient use of spectrum is another design aspect with sustainability implications. \ac{TN} typically achieve high spectral efficiency through frequency reuse and localized coverage. In \ac{NTN}, however, the presence of very large cells means that spatial frequency reuse is inherently limited \cite{10355089}. A single satellite beam can cover an area that would contain dozens of terrestrial cells, yet it has to share a finite bandwidth among all those users. This disparity can lead to lower per-user throughput and spectral efficiency in \ac{NTN} compared to \ac{TN}. Moreover, satellites often operate in comparatively lower frequency bands (e.g., L, S, or C bands for broad coverage) which have limited bandwidth, or in higher bands (Ka/Q/V) that suffer more propagation loss.


From a Net-Zero perspective, spectral efficiency links to \ac{EE}: carrying more data per Hz means that the network can meet demand with fewer transmissions, saving power. Thus, integrated \ac{TN} and \ac{NTN} must strive for spectral-efficiency maximization under the unique constraints of \ac{NTN}. This includes developing new air interface adaptations for long delays and Doppler (to reduce overhead), efficient multicast/broadcast in satellites (one-to-many transmission can save spectrum and energy for common content delivery), and perhaps novel multi-access schemes tailored for mixed integrated \ac{TN} and \ac{NTN} environments.

\begin{figure*}[t!]
	\centering
	\includegraphics[scale=0.355]{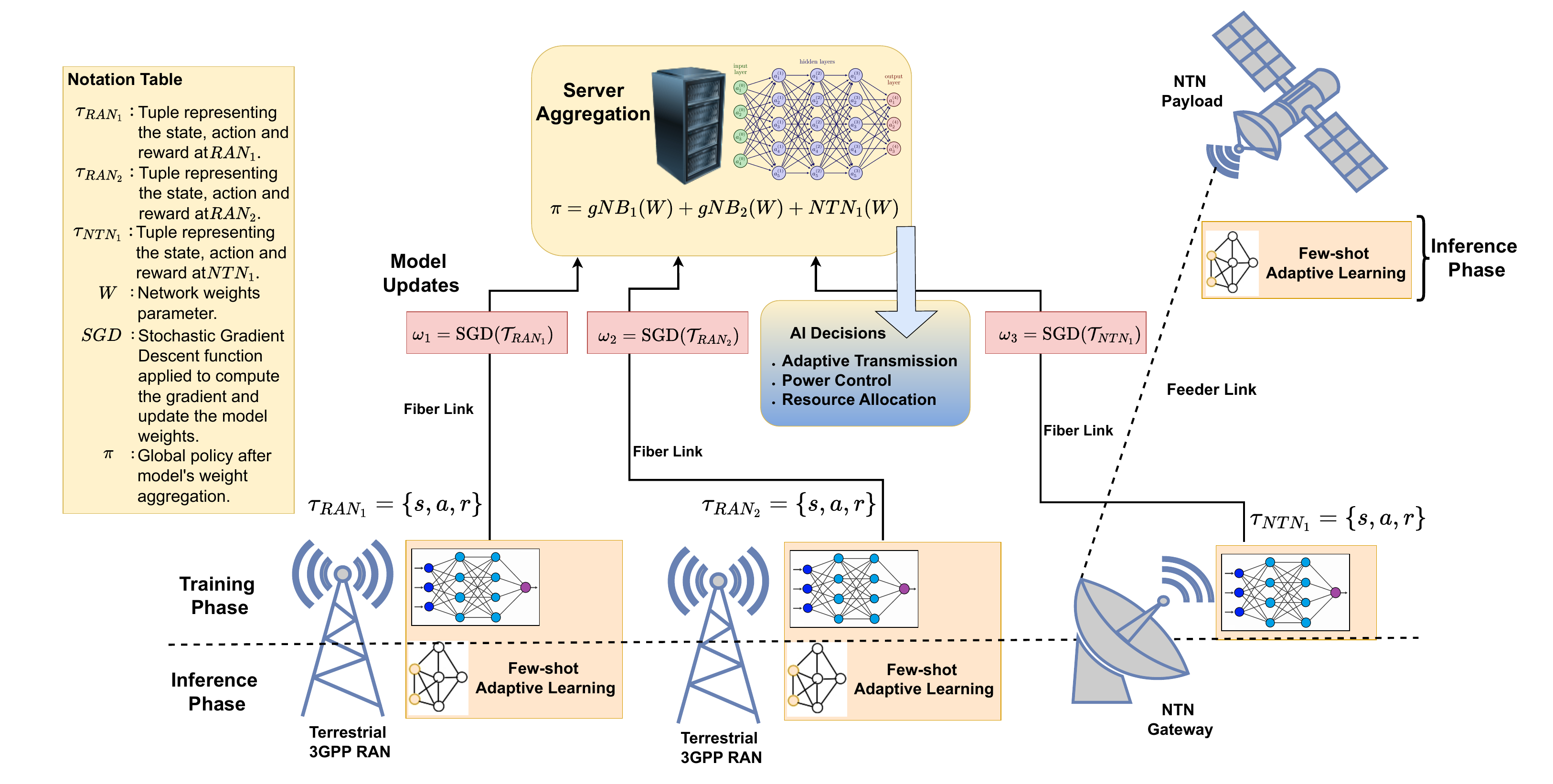}
	\caption{Energy efficient few-shot FL framework for proactive resource orchestration.}
	\label{fige1}
\end{figure*}

\section{Key Enablers of Net-Zero integrated TN and NTN}
Achieving Net-zero emissions in integrated \ac{TN} and \ac{NTN} requires a multi-faceted approach that synergizes renewable energy adoption, intelligent energy management, efficient network design, and sustainability-oriented policy frameworks. This section outlines the primary technological and operational enablers facilitating the transition toward carbon-neutral \ac{TN} and \ac{NTN} deployments.

\subsection{Renewable and Sustainable Energy Integration}

The decarbonization of \ac{TN} and \ac{NTN} infrastructure depends on large-scale integration of renewable energy sources across both terrestrial and non-terrestrial segments. Solar-powered aerial and space platforms, including \ac{HAPS}, \ac{UAV}, and \ac{LEO} satellites, now feature lightweight, high-efficiency multi-junction photovoltaic cells that convert solar radiation into electrical power. This enables persistent flight and long-term orbital operations without frequent refueling or battery replacement. Such approaches reduce reliance on ground-based fossil-fuel energy sources, aligning with the \ac{3GPP} TR 38.821 discussions on renewable-powered \ac{NTN} gateways \cite{tr1}. Commercial deployments, such as Starlink’s photovoltaic-optimized satellites and Airbus Zephyr’s solar \ac{HAPS}, exemplify the shift toward Net-Zero energy strategies in integrated \ac{TN} and \ac{NTN} environments.

In parallel, green-powered ground infrastructure is emerging as a key enabler of sustainable \ac{TN} and \ac{NTN} operations. Next-generation terrestrial \ac{BS}, relay nodes, and \ac{NTN} gateways are being designed to operate on renewable sources such as solar, wind, or hybrid systems. Interconnection through smart microgrids enables bidirectional energy flow and dynamic load balancing, ensuring service continuity. Integrating such renewable-powered infrastructure into terrestrial and non-terrestrial backhaul reduces the carbon footprint while enhancing resilience during outages and disaster recovery. \ac{3GPP} TR 38.821 \cite{tr1} highlights the strategic role of renewable-powered \ac{NTN} gateways in meeting Net-Zero energy targets for integrated \ac{TN} and \ac{NTN} deployments.

Additionally, ambient energy harvesting offers a pathway for powering low-power devices and auxiliary subsystems with minimal reliance on stored or grid-supplied energy. Techniques such as rectenna-based \ac{RF} energy harvesting arrays, alongside kinetic and thermal energy recovery mechanisms, capture energy from the surrounding environment \cite{ar2}. \ac{RF} harvesting, for instance, converts ambient \ac{RF} emissions from \ac{TN}, satellite downlinks, or broadcast systems into usable DC power, supporting the operation of sensors and control modules within the integrated \ac{TN} and \ac{NTN} ecosystem.

\subsection{Artificial Intelligence-Driven Energy Management}
\ac{AI} is pivotal for optimizing energy usage in heterogeneous \ac{TN} and \ac{NTN} environments. \ac{FL} enables distributed training at terrestrial and non-terrestrial nodes, supporting energy-aware optimization without transferring raw data to centralized servers \cite{10729847}. This preserves privacy, reduces backhaul overhead, and lowers energy costs. By enabling near real-time adaptation of scheduling, resource allocation, and power control, \ac{FL} fosters self-optimizing networks aligned with Net-Zero carbon targets.

\ac{ML}–driven predictive resource orchestration frameworks further enhance \ac{EE} by leveraging historical and real-time datasets to forecast traffic surges, renewable energy generation patterns, and mobility trends across both \ac{TN} and \ac{NTN} \cite{ar3}. Fig. \ref{fige1} shows the energy-efficient few-shot \ac{FL} framework for proactive resource orchestration. These forecasts enable dynamic satellite beam reconfiguration, \ac{UAV} repositioning for coverage and solar exposure, and adaptive scheduling to balance service quality with energy conservation.

Standardization efforts in \ac{3GPP} SA2 and European Telecommunication Standard Association (ETSI) Experiential Networked Intelligence (ENI) are advancing \ac{AI}-driven green networking frameworks for integrated \ac{TN} and \ac{NTN} deployments. They emphasize energy-centric KPI for \ac{NTN} links, such as renewable energy utilization, and carbon intensity per bit enabling interoperable and energy-aware orchestration aligned with global sustainability goals \cite{ar4}.
\begin{figure*}[ht!]
	\centering
	\includegraphics[width=0.90\linewidth]{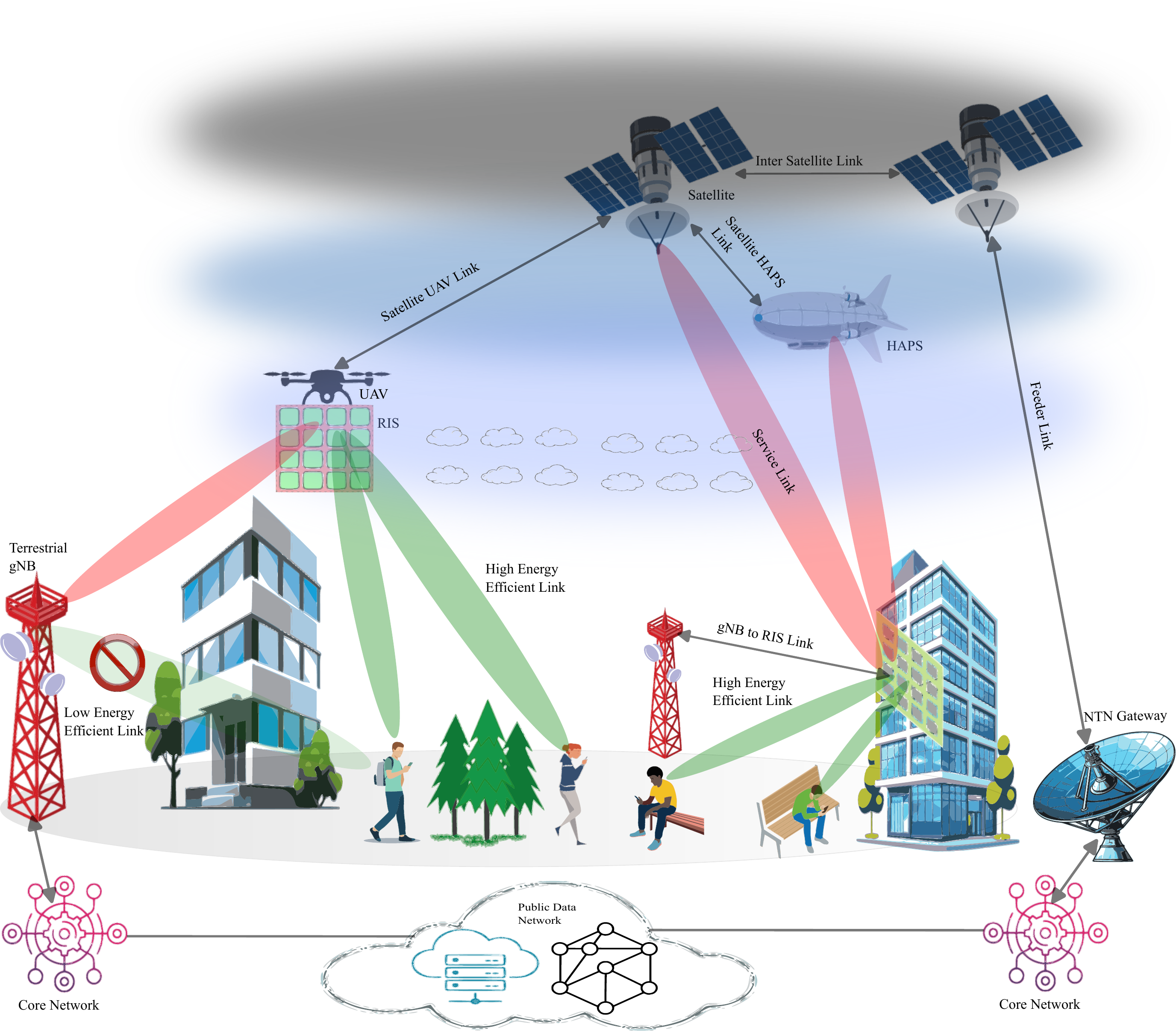}
	\caption{RIS-aided steered beam for downlink communication.}
	\label{fige11}
\end{figure*}

\subsection{Energy-Efficient Network Design and Protocols}
Minimizing operational energy consumption requires network architectures and protocol-level optimizations. Low-Power Wake-Up Signals (LP-WUS) enable \ac{NTN}-\ac{IoT} devices to remain in ultra-low-power states until activated by lightweight control signals. This significantly reduces idle power drain, especially in massive \ac{IoT} deployments. This approach aligns with \ac{3GPP} Release 19 enhancements for extended Discontinuous Reception (eDRX) and Power Saving Mode (PSM) \cite{9356526}.

Cell-free, user-centric architectures further contribute to \ac{EE} by coordinating disaggregated \ac{AP} through a centralized processing unit. By eliminating cell boundaries and dynamically serving users based on channel conditions, they reduce redundant transmissions and improve \ac{EE} per bit. In \ac{TN} and \ac{NTN} integration, distributed \ac{AP} clusters across terrestrial nodes, \ac{UAV}, and \ac{LEO} satellites can jointly serve users, enabling seamless mobility and adaptive power allocation in line with green networking objectives.

\begin{figure*}[t!]
    \centering
    \includegraphics[width=0.85\linewidth]{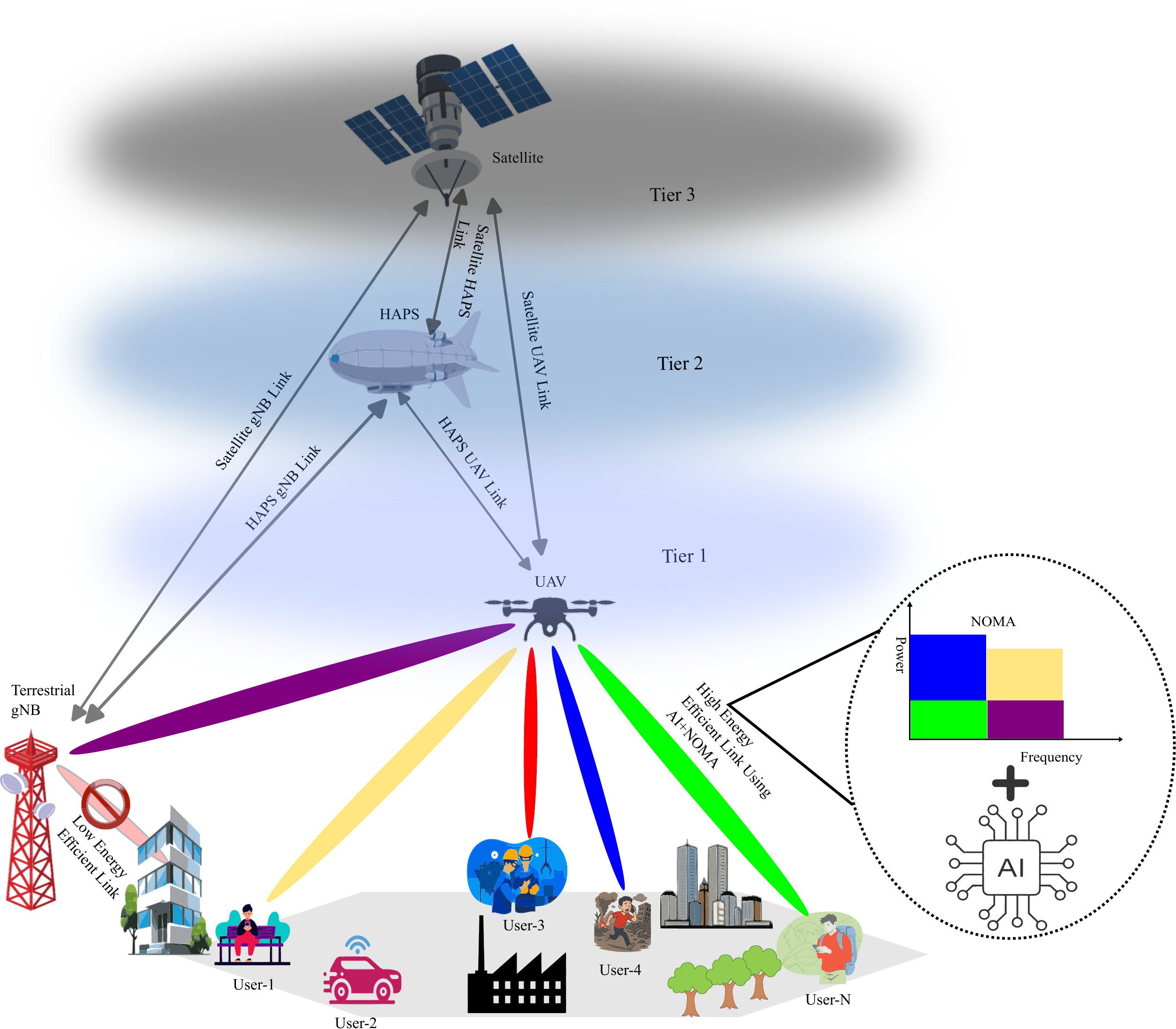}
    \caption{Illustration of integrated \ac{TN} and \ac{NTN} system model for various usage scenarios.}
    \label{fig3}
\end{figure*}
\subsection{Reconfigurable and Passive Energy-Saving Technologies}

Passive RIS elements can significantly lower the energy footprint of signal propagation in \ac{TN} and \ac{NTN} systems \cite{bariah2022ris}. RIS integrated on terrestrial \ac{BS}, building facades, or aerial platforms use programmable meta-material elements to passively reflect, refract, or focus incident signals toward intended receivers. Fig. \ref{fige11} shows the RIS-aided steered beam for downlink communication. Unlike traditional active repeaters, RIS operates without \ac{RF} amplification, instead relying on precise phase-shift control to manipulate wavefronts. This approach reduces transmission power requirements, minimizes interference, and improves spectral efficiency.

Ongoing work in \ac{3GPP} \ac{RAN}1 and \ac{RAN}2 is also exploring the integration of RIS into \ac{NTN} payloads to improve \ac{EE} in space-to-ground communications. Embedding RIS panels on satellites or \ac{HAPS} enables reflected or refracted beams to be dynamically shaped, improving link budgets while minimizing onboard \ac{RF} power amplification. Such RIS-assisted \ac{NTN} links represent a promising direction for reducing operational energy demand while maintaining robust connectivity across integrated \ac{TN} and \ac{NTN} architectures.

\subsection{Policy, Standardization, and Carbon Accountability}
Policy and standardization efforts provide the governance structure necessary for sustainable \ac{TN} and \ac{NTN} operations. Standards bodies such as \ac{ITU}, \ac{3GPP}, and IEEE are actively defining KPI that focus on \ac{EE}, renewable energy integration, and carbon neutrality. These KPI establish measurable benchmarks for system-level sustainability, including energy consumed per delivered bit, renewable penetration rate, and lifecycle $CO_2$ emissions. Embedding such metrics into global standards enables consistent cross-vendor evaluation, facilitates compliance reporting, and accelerates the adoption of green networking practices across terrestrial, aerial, and satellite domains.


\section{Case Study}
In this section, we have proposed an \ac{AI}-based algorithm to enhance the \ac{EE} of users (in different application scenarios) in integrated \ac{TN} and \ac{NTN}, having varying data rate requirements, as illustrated in Fig.~\ref{fig3}. Following the approach in \cite{jamshed2021unsupervised}, \ac{NOMA} is used to perform subcarrier allocation and unsupervised learning is applied for user clustering. However, instead of relying on K-means or K-medoids, we employ spectral clustering. Unlike traditional clustering methods that assume spherical and equally sized clusters, spectral clustering constructs a similarity graph and leverages eigenvector-based embeddings. This allows it to identify clusters of irregular shape, size, and density, which makes it more suitable for large-scale and unevenly distributed data. We consider an uplink transmission scenario within a coverage radius of 500 meters. Since the proposed algorithm is aimed at improving the \ac{EE} of the users performing uplink communication, our simulations are performed up to Tier 1, as shown in Fig.~\ref{fig3}. The number of subcarriers is fixed at 128, with an uplink transmit power of 0.2 W, while the circuit power is set to 5 dBm. Since direct communication with the \ac{BS} is obstructed, a \ac{UAV} at Tier 1, positioned at an altitude of 2000 meters, serves as a relay to maintain connectivity. The optimization problem formulated here is non-convex. To address this, we first perform subcarrier allocation under fixed power conditions. User grouping is carried out using spectral clustering, and subcarriers are assigned based on mean channel gain. Subsequently, power allocation is refined using iterative optimization techniques.

\begin{figure}[t!]
    \centering
    \includegraphics[width=\columnwidth]{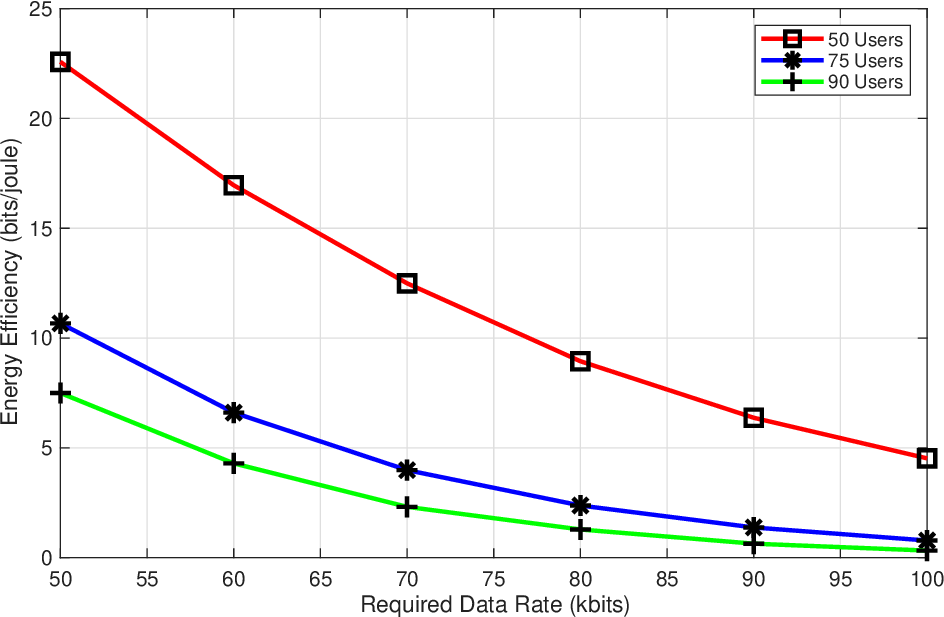}
    \caption{\ac{EE} versus varying required data rate for different values of fixed number of users.}
    \label{fig4}
\end{figure}

Fig.~\ref{fig4} presents the performance comparison of the proposed algorithm across different user densities and data rate requirements. The results show that as the number of devices increases and higher data rates are demanded, the \ac{EE} of the system decreases, following an exponential trend. Overall, the integration of \ac{UAV} with \ac{NOMA} and spectral clustering demonstrates significant potential in improving system-wide \ac{EE} while supporting Net-Zero emissions. Although spectral clustering introduces higher computational complexity compared to simpler clustering methods, the resulting improvements in cluster quality and resource allocation justify this trade-off in energy-constrained communication systems.

\begin{figure}[t!]
    \centering
    \includegraphics[width=\linewidth]{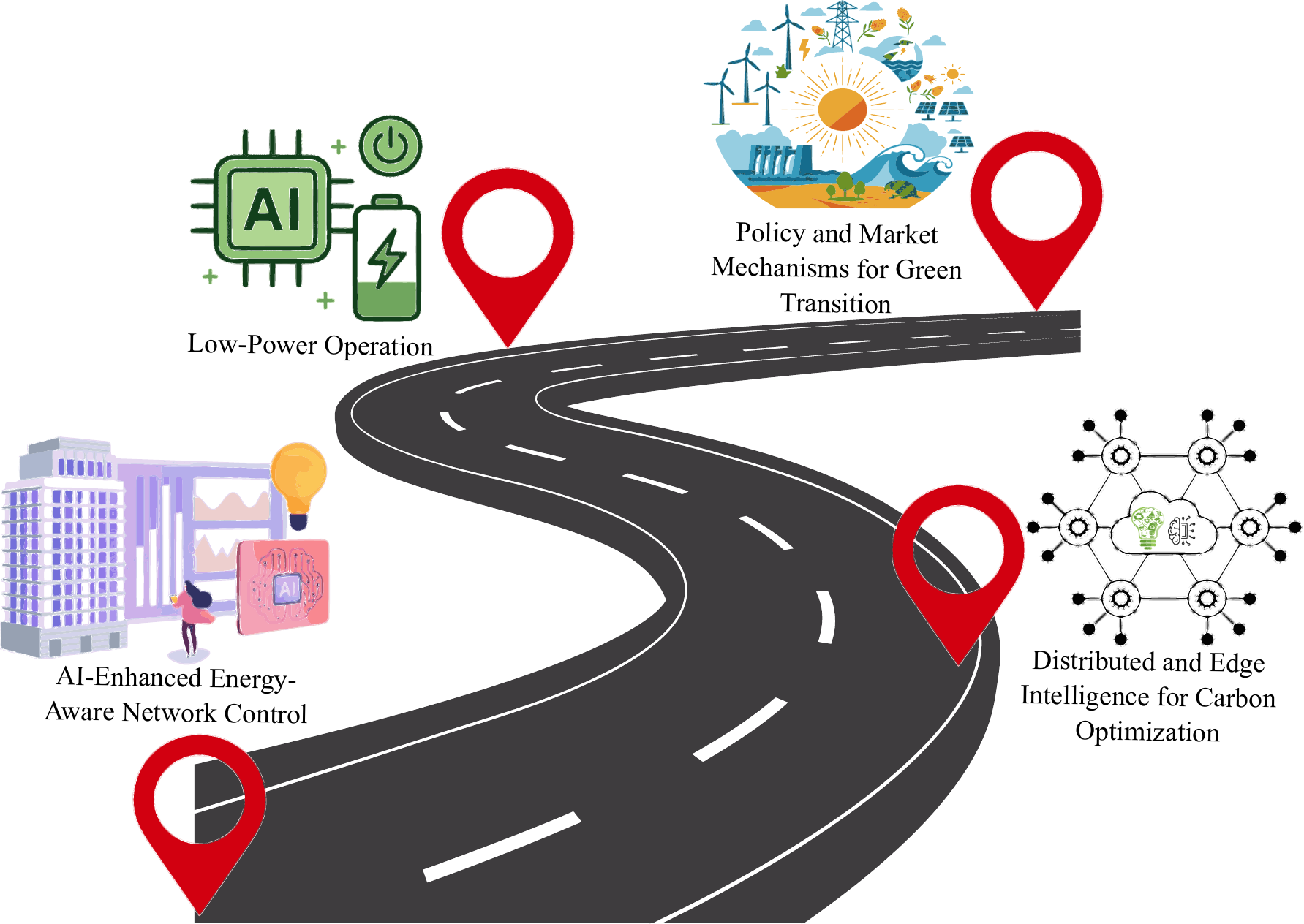}
    \caption{An overview of future road map to achieve Net-Zero targets in integrated TN and NTN.}
    \label{fig4a}
\end{figure}
\section{Future Directions}
While current key enablers provide the foundation for low-carbon \ac{TN} and \ac{NTN} deployments, achieving full Net-Zero compliance will require innovations that go beyond today’s technical maturity. This section, as shown in Fig. \ref{fig4a}, outlines research trajectories, emerging technologies, and evolving standardization efforts that could shape the sustainable evolution of integrated \ac{TN} and \ac{NTN}.

\subsection{Artificial Intelligence-Enhanced Energy-Aware Network Control}
Integrating multi-layer \ac{AI} models with renewable energy forecasts enables \ac{TN} and \ac{NTN} orchestration systems to perform real-time, energy-aware routing and workload placement across terrestrial, aerial, and orbital nodes. \ac{AI}-driven orchestration dynamically redistributes computational tasks and reroutes traffic toward nodes with higher renewable energy availability or lower carbon intensity, reducing reliance on carbon-intensive backup power.  

In parallel, when combined with \ac{AI}-driven beam management, RIS can adapt in real time to user mobility and environmental dynamics, lowering transmission energy costs in space-to-ground links. Achieving this requires coordination across the physical, medium access control, and network layers to enable synchronized reconfiguration of multiple RIS-enabled \ac{TN} and \ac{NTN} nodes. This cross-layer orchestration aligns beam steering with link quality and \ac{EE} targets, ensuring that coverage adjustments minimize redundant transmissions and prioritize nodes powered by renewable sources, thus maintaining low-carbon and high-performance connectivity.

\subsection{Distributed and Edge Intelligence for Carbon Optimization}
Deploying lightweight \ac{AI} inference engines directly on satellites, \ac{UAV}, and edge \ac{BS} enables real-time decision-making without constant reliance on cloud or core network connectivity. This reduces long-haul backhaul transmissions, minimizes latency for mission-critical \ac{IoT} and \ac{NTN} services, and lowers the energy consumption associated with large-scale data transport. Techniques such as TinyML, model quantization, and pruning make it feasible to run \ac{AI} models efficiently on resource-constrained platforms while preserving high inference accuracy. 

Moreover, \ac{AI}-assisted predictive caching can be extended to multi-orbit content placement strategies, where data is pre-positioned not only at terrestrial edge nodes but also within \ac{LEO} or \ac{MEO} payload storage \cite{11049853}. Such localization reduces repetitive high-power satellite downlinks and enables energy-proportional content delivery. Furthermore, adaptive task offloading based on renewable energy availability improves latency and \ac{EE}. \ac{AI}-driven context-aware algorithms dynamically distribute tasks across satellites, \ac{UAV}, edge servers, and terrestrial nodes to reduce carbon footprint, extend device lifespan, and preserve \ac{QoS}.


\subsection{Low-Power Operation Beyond 3GPP Release 19}
Extending LP-WUS with \ac{AI}-driven activation prediction and context-aware sleep scheduling can significantly enhance \ac{NTN}-\ac{IoT} \ac{EE} without incurring latency penalties. By analyzing historical traffic patterns, device mobility profiles and environmental conditions, \ac{AI} models can predict optimal wake-up intervals and preemptively activate nodes only when meaningful data exchange is likely. This predictive approach minimizes unnecessary signaling while ensuring timely response.

Context-aware sleep cycles can be dynamically adapted according to application-specific latency requirements and real-time renewable energy availability, allowing idle devices to remain in ultra-low-power states for extended periods. Achieving this vision will require coordinated enhancements to \ac{RAN}–core signaling to streamline wake-up procedures, reduce control plane overhead, and maintain tight synchronization between \ac{TN} and \ac{NTN} segments \cite{7949048}. 



\subsection{Policy and Market Mechanisms for Green Transition}

Dynamic regulatory frameworks for the green transition in integrated \ac{TN} and \ac{NTN} ecosystems can integrate performance-linked incentives such as subsidies, tax benefits or preferential spectrum allocation, directly tied to measurable sustainability outcomes. \ac{AI}-enabled monitoring systems, combined with blockchain-backed audit trails, can provide continuous, tamper-proof compliance verification, ensuring transparency in meeting environmental commitments. By linking economic rewards to real-time \ac{EE} and low-carbon operational performance, these mechanisms would encourage operators to adopt renewable-powered infrastructure, as well as optimize manufacturing and operational processes. Such market-aligned policy structures could accelerate industry-wide decarbonization while fostering innovation across terrestrial, aerial, and satellite network domains.









\section{Conclusion} 
In this article, we presented a comprehensive vision for achieving Net-Zero targets in integrated \ac{TN} and \ac{NTN}. We highlighted the unique design challenges arising from combining heterogeneous ground, aerial, and space systems, and identified key enablers such as renewable-powered infrastructure, \ac{AI}-driven energy management, and RIS. A case study demonstrated that integrating \ac{UAV}, \ac{AI}, and \ac{NOMA} can significantly enhance \ac{EE} in various deployment scenarios, even under high user density and data rate demands. The integration of distributed intelligence, low-power operation, and sustainability-oriented policy frameworks is essential to ensure that 6G and beyond networks deliver ubiquitous connectivity while minimizing the carbon impact. By incorporating Net-Zero design principles and standardization from the outset, integrated \ac{TN} and \ac{NTN} systems can play a fundamental role in enabling universal connectivity and long-term environmental sustainability.

\bibliographystyle{IEEEtran}

\bibliography{IEEEabrv,main}

\end{document}